\documentclass[twocolumn,showpacs,prl]{revtex4}
\usepackage{epsfig}
\usepackage{bm}
\usepackage{graphicx}
\newcommand{\vnabla}{{\mbox{\boldmath$\nabla$}}}

\newcommand{\vR}{{\mbox{\boldmath$R$}}}
\newcommand{\vD}{{\mbox{\boldmath$D$}}}

\newcommand{\vA}{{\mbox{\boldmath$A$}}}
\newcommand{\va}{{\mbox{\boldmath$a$}}}
\newcommand{\br}{{\mbox{\boldmath$r$}}}
\newcommand{\vG}{{\mbox{\boldmath$G$}}}
\newcommand{\vQ}{{\mbox{\boldmath$Q$}}}

\newcommand{\vT}{{\mbox{\boldmath$T$}}}

\newcommand{\vk}{{\mbox{\boldmath$k$}}}

\newcommand{\vb}{{\mbox{\boldmath$b$}}}
\newcommand{\vB}{{\mbox{\boldmath$B$}}}

\newcommand{\vtau}{{\mbox{\boldmath$\tau$}}}

\newcommand{\vd}{\mbox{\boldmath$d$}}

\begin{document}

\title{Order parameter and vortices in the superconducting $Q$-phase of CeCoIn$_5$}

\author{D.F. Agterberg$^1$, M. Sigrist$^2$, and H. Tsunetsugu$^3$}
\address{$^1$ Department of Physics, University of Wisconsin-Milwaukee, Milwaukee, WI 53211}
\address{$^2$ Theoretische Physik ETH-H\"onggerberg CH-8093 Z\"urich, Switzerland}
\address{$^3$ Institute for Solid State Physics, University of Tokyo, Kashiwa, Chiba 277-8581, Japan}


\begin{abstract}
Recently, it has been reported that the low-temperature
high-magnetic field phase in CeCoIn$_5$ ($Q$-phase), has
spin-density wave (SDW) order that only exists within this phase.
This indicates that the SDW order is the result of the development
of pair density wave (PDW) order in the superconducting phase that
coexists with $d$-wave superconductivity. Here we develop a
phenomenological theory for these coexisting orders. This provides
selection rules for the PDW order and further shows that the
detailed structure of this order is highly constrained. We then
apply our theory to the the vortex phase. This reveals vortex
phases in which the $d$-wave vortex cores exhibit charge density
wave (CDW) order and further reveals that the SDW order provides a
unique probe of the vortex phase.

\end{abstract} \maketitle

The low-temperature high-magnetic field phase in CeCoIn$_5$
($Q$-phase) has been thought to be the best example of a FFLO
superconductor \cite{bia03,rad03,ful64,lar65} and has thus
generated a tremendous interest \cite{cas04,mat07}. However, the
recent measurements of Kenzelmann {\it et al.} \cite{ken08},
suggest that this point of view should be altered. This important
experimental discovery shows that the $Q$-phase reveals itself
through the appearance of an incommensurate spin density wave
(SDW) order. What is striking about this SDW order is that it
vanishes when superconductivity vanishes at high magnetic fields.
This implies that superconducting order is the primary order
parameter with the SDW order induced as a secondary order
parameter. A possibility for such superconducting order, as
pointed out by Kenzelmann {\it et al.} \cite{ken08}, is pair
density wave (PDW) superconductivity. If the SDW order is
associated with a wavevector $\vQ$, then the PDW order must have
the wavevector $-\vQ$ to be able to induce the SDW order. The SDW
order has $\vQ=(q,q,0.5)$, which is too large to be a consequence
of the long-wavelength modulation of a FFLO phase
\cite{ful64,lar65,miy08}. The PDW order is more akin to the
$\pi$-triplet staggered pairing suggested by Aperis \cite{ape08}
or to the the PDW order suggested in La$_{2-x}$Ba$_x$CuO$_4$ at
$x=1/8$ \cite{ber07}. The ensuing physical picture is then a
$d$-wave superconductor at low fields with PDW order appearing
through a second order phase transition at high fields. These two
types of superconducting order will coexist in the $Q$-phase.

The observation of this PDW order raises a series of deep
questions about the origin of this phase. To help address these,
we have developed a phenomenological theory for this PDW order.
Our approach complements that given by Kenzelmann {\it et al.} and
is based on irreducible representations of the full space group.
We find that this theory strongly constrains the PDW order and
provides useful information about the vortex phase. \\

\begin{figure}
\epsfxsize=1.5 in \center{\epsfbox{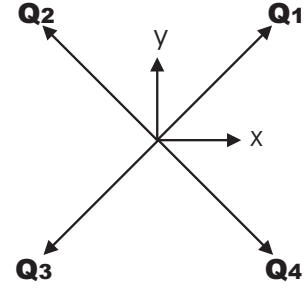}}
\caption{Directions of $\vQ_i$ used in the text. The field is
applied along the direction $\vQ_1$}. \label{fig1}
\end{figure}

{\it PDW superconducting order parameter:} Our approach is to
classify the PDW order in terms of irreducible representations of
the full space group \cite{bou36}. For CeCoIn$_5$ this is
$P4/mmm$. For order appearing at a wavevector $\vQ$, the order
parameter is defined by the irreducible representations of $G_Q$
(set of elements conserving $\vQ$) and the star of the wavevector
$\vQ$ (set of wavevectors symmetrically equivalent to $\vQ$). For
$\vQ=(q,q,0.5)=(q,q,-0.5)$
$G_Q=\{E,C_{2\eta},\sigma_z,\sigma_{\zeta}\}$ with $ C_{2 \eta} $
the $180^{o} $-rotation around the axis $(1,1,0) $, $ \sigma_z $
and $ \sigma_{\zeta} $ the mirror operations at the basal plane
and the plane perpendicular to $(1,-1,0) $, respectively. Note $
(0,0,1) $ is a reciprocal lattice vector. In Table I, we give the
irreducible representations of $G_Q$  together with representative
basis functions for spin-singlet pairing (scalar functions
$\psi(\vk)$ \cite{sig91}), spin-triplet pairing (vector functions
$\vd(\vk)$ \cite{sig91}), and spin density order ($S_i$). To
define the additional order parameter components at the
wavevectors in the star of $\vQ$ we use the elements
$\{E,C_4,C_4^2,C_4^3\}$ (these give the star of $ \vQ$,
$\{\vQ_1,\vQ_2,\vQ_3,\vQ_4\}$ respectively, as shown in Fig.~1).
This then defines a superconducting order parameter with four
components which we define as $ \Delta_{\Gamma_i} =
(\Delta_{\Gamma_i,Q_1},\Delta_{\Gamma_i,Q_2},\Delta_{\Gamma_i,Q_3},\Delta_{\Gamma_i,Q_4})$.
With these definitions, the symmetry properties of the order
parameter are given as follows ($D_{\Gamma_i}(g)$
defined in Table I):  translation $\vT$,
$\Delta_{\Gamma_i,Q_j}\rightarrow
e^{i\vQ_j\cdot\vT}\Delta_{\Gamma_i,Q_j}$ (
$\Delta_{\Gamma_i,Q_j}^*\rightarrow
e^{-i\vQ_j\cdot\vT}\Delta_{\Gamma_i,Q_j}^*$); time-reversal operation
$\Delta_{\Gamma_i,Q_j}\rightarrow
\Delta_{\Gamma_i,-Q_j}^*$. Moreover, the transformations $ G_Q $
lead to
\begin{equation} \nonumber \begin{array}{ll}
C_4: & D_{\Gamma_i} (C_4) (\Delta_{\Gamma_i,Q_2},\Delta_{\Gamma_i,Q_3},\Delta_{\Gamma_i,Q_4},\Delta_{\Gamma_i,Q_1}) \\
\sigma_z: & D_{\Gamma_i}(\sigma_z)(\Delta_{\Gamma_i,Q_1},\Delta_{\Gamma_i,Q_2},\Delta_{\Gamma_i,Q_3},\Delta_{\Gamma_i,Q_4}) \\
C_{2\eta}: & D_{\Gamma_i}(C_{2\eta})(\Delta_{\Gamma_i,Q_1},\Delta_{\Gamma_i,Q_4},\Delta_{\Gamma_i,Q_3},\Delta_{\Gamma_i,Q_2}) \\
\sigma_{\zeta} : & D_{\Gamma_i}(\sigma_{\zeta})
(\Delta_{\Gamma_i,Q_1},\Delta_{\Gamma_i,Q_4},\Delta_{\Gamma_i,Q_3},\Delta_{\Gamma_i,Q_2})
\end{array}
\end{equation}



Table~\ref{tab1} reveals that both singlet and triplet order
parameters belong to the same representation which implies that
singlet and triplet superconductivity are mixed. This mixing is
due to spin-orbit coupling which cannot be justifiably ignored in
CeCoIn$_5$. Formally, this is a consequence of the fact that $G_Q$
does not contain inversion symmetry \cite{fri04}. Previous studies
have examined the development of singlet-triplet mixing in related
situations \cite{min94,shi00,leb06,kab07,ken08}, often through
Lifshitz invariants that appear in the Ginzburg Landau free energy
when parity symmetry is broken. We have confirmed that our
formalism yields the same results as through the use of Lifshits
invariants.

\begin{widetext}
\begin{table}
\begin{tabular}{|c|c|c|c|c|c|c|c|}
  \hline
  Irrep ($\Gamma_i$)& $D_{\Gamma_i}(E)$ & $D_{\Gamma_i}(\sigma_z)$ & $D_{\Gamma_i}(C_{2\eta})$ & $D_{\Gamma_i}(\sigma_{\zeta})$ & Representative $\psi(\vk)$)&  Representative $\vd(\vk)$& Representative $S_i$ \\
  \hline
  $\Gamma_1$& 1 & 1 & 1 & 1 & $s$, $k_xk_y$ & $\hat{z}(k_x-k_y)$, $k_z(\hat{x}-\hat{y})$ & \\
  $\Gamma_2$ & 1 & 1 & -1 & -1 & $k_x^2-k_y^2$& $\hat{z}(k_x+k_y)$, $k_z(\hat{x}+\hat{y})$& $S_z$\\
  $\Gamma_3$ & 1 & -1 & -1 & 1 & $k_z(k_x+k_y)$ & $\hat{x}k_x-\hat{y}k_y$,$\hat{x}k_y-\hat{y}k_x$& $S_x-S_y$\\
  $\Gamma_4$ & 1 & -1 & 1 & -1 & $k_z(k_x-k_y)$& $\hat{x}k_x+\hat{y}k_y$,$\hat{x}k_y+\hat{y}k_x$& $S_x+S_y$ \\
  \hline
\end{tabular}
\caption{ Representative spin-singlet, spin-triplet, and spin
density basis functions for the different irreducible
representations that have momentum $\vQ_1=(q,q,0.5)$.}
\label{tab1}
\end{table}
\end{widetext}

{\it Free Energy and PDW solutions:} We use a Ginzburg Landau
theory to describe the PDW and $d$-wave order parameters. While
this will not be reliable on a quantitative level, it will allow
us to correctly identify the properties of the PDW order and make
robust experimental predictions. The PDW Ginzburg Landau free
energy density is constructed by imposing invariance under the
above symmetries (note that this free energy density is the same
for all $\Gamma_l$),
\begin{widetext}
\begin{equation}\begin{array}{ll}
f=&\alpha\sum_{i}|\Delta_{\Gamma_l,Q_i}|^2+\beta_1(\sum_{i}|\Delta_{\Gamma_l,Q_i}|^2)^2
+
\beta_2\sum_{i<j}|\Delta_{\Gamma_l,Q_i}|^2|\Delta_{\Gamma_l,Q_j}|^2+\beta_3(|\Delta_{\Gamma_l,Q_1}|^2|\Delta_{\Gamma_l,Q_3}|^2
+|\Delta_{\Gamma_l,Q_2}|^2|\Delta_{\Gamma_l,Q_4}|^2) \\
&+\beta_4[\Delta_{\Gamma_l,Q_1}\Delta_{\Gamma_l,Q_3}(\Delta_{\Gamma_l,Q_2}\Delta_{\Gamma_l,Q_4})^*+
(\Delta_{\Gamma_l,Q_1}\Delta_{\Gamma_l,Q_3})^*\Delta_{\Gamma_l,Q_2}\Delta_{\Gamma_l,Q_4}]
 \\
&+\kappa_1\sum_i|\vD \Delta_{\Gamma_l,Q_i}|^2+ \kappa_2\sum_i
(-1)^i(
|D_1\Delta_{\Gamma_l,Q_i}|^2-|D_2\Delta_{\Gamma_l,Q_i}|^2)+
\kappa_3\sum_i|D_z\Delta_{\Gamma_l,Q_i}|^2
+\frac{1}{2}(\nabla\times\vA)^2
\end{array}
\label{free}
\end{equation}
\end{widetext}
where $\vD=-i\vnabla- 2e \vA$, $\vB=\nabla\times\vA$,  $D_1$
corresponds to the $(1,1,0)$, and $D_2$  to the $(1,-1,0)$
direction. The free energy density for the $d$-wave order
parameter is taken to be:
\begin{equation}
f_d=\alpha_d|\Delta_d|^2+\beta_d|\Delta_d|^4+\kappa|\vD\Delta_d|^2+\kappa_c|D_z\Delta_d|^2
\label{d-free}
\end{equation}
The coupling between these order parameters is given by (note that
this coupling is the same for all $\Gamma_l$):
\begin{equation}\begin{array}{ll}
f_c = & \beta_{c1} \sum_i |\Delta_d|^2|\Delta_{\Gamma_l,Q_i}|^2  \\
& +
\beta_{c2}[\Delta_d^2(\Delta_{\Gamma_l,Q_1}\Delta_{\Gamma_l,Q_3}+\Delta_{\Gamma_l,Q_2}\Delta_{\Gamma_l,Q_4})^* +c.c.] \nonumber
\end{array}
\label{PDWdc}
\end{equation}

This free energy is similar to one studied earlier in the context
of PDW order in  Ref.~\cite{agt08}. The ''homogeneous'' phase in
the absence of a magnetic field has five PDW states distinct by
symmetry, if we ignore the $d$-wave phase (the phase factors
$\phi_1, \phi_2,$ and $\phi_3$ are not determined by the free
energy):
\begin{equation} \begin{array}{l}
\Delta_{\Gamma_l}^{(1)} = (e^{i\phi_1},0,0,0)  \\
\Delta_{\Gamma_l}^{(2)} = (e^{i\phi_1},e^{i\phi_2},0,0) \\
\Delta_{\Gamma_l}^{(3)} = (e^{i\phi_1},0,e^{i\phi_3},0) \\
\Delta_{\Gamma_l}^{(4)} =
(e^{i\phi_1},e^{i\phi_2},e^{i\phi_3},e^{i(\phi_1+\phi_3-\phi_2)})\\
 \Delta_{\Gamma_l}^{(5)} =
(e^{i\phi_1},ie^{i\phi_2},e^{i\phi_3},ie^{i(\phi_1+\phi_3-\phi_2)}).
\end{array}
\end{equation}

This set is reduced to  $ \Delta_{\Gamma_l}^{(3)} $ and $
\Delta_{\Gamma_l}^{(4)} $ in the presence of a $d$-wave order
parameter \cite{agt08}. Finally, a magnetic field in the basal
plane along the $(1,1,0$)-direction favors the state $
\Delta_{\Gamma_l}^{(3)} $ as it removes the degeneracy between the
$\vQ_1$ and $\vQ_2$ wavevectors (the pairs $\vQ_1,\vQ_3$ and
$\vQ_2,\vQ_4$ remain degenerate). As an example we assume that the
field lies along $ (1,-1,0) $-direction and yields the state $
(0,e^{i\phi_2}, 0 , e^{i\phi_4}) $. Choosing $ \phi_2 = \phi_4=0$,
the spatial dependence of the PDW order is given by
$\Delta_{\Gamma_l,Q_2}\cos(\vQ_2\cdot \vR)$. In view of the
coupling to the $d$-wave order parameter the relative phase
between $\Delta_d$ and $\Delta_{\Gamma_l,Q_2}$ can be either $0 $
($\pi$) or $ \pm \pi/2 $ \cite{agt08} which are both permitted by
the free energy. Generally, the combined PDW and $d$-wave
superconductivity must then take one of two forms when vortices
are ignored: $\Delta_d+i|\Delta_{\Gamma_l,2}|\cos(\vQ_2\cdot\vR)$
(time reversal violating phase) or
$\Delta_d+|\Delta_{\Gamma_l,2}|\cos(\vQ_2\cdot\vR)$ (time-reversal
invariant phase).

{\it Coupling to spin-density wave:} The SDW order can be induced
through the combined PDW and $d$-wave superconductivity
\cite{ken08}. We assume here that the SDW is sufficiently weak so
as to not alter the free energy significantly. The free energy
density for the SDW order is then simply
$ f_{SDW}=\alpha_s S^z_{Q_2}S^z_{-Q_2}+f_{coupling} $
with $\alpha_s>0$. To determine $f_{coupling}$, it is important to
note that the SDW order breaks time reversal and can be generated
by the PDW and $d$-wave order in two ways. The first is by
coupling directly to the time-reversal symmetry violating phase
and the second is by coupling to the applied magnetic field and
the time-reversal invariant phase. This leads to two possible
coupling terms that appear in $f_{coupling}$, the first exists
without the magnetic field,
\begin{equation}
\gamma_1 iS^z_{Q_2} \{ \Delta_{d}^*\Delta_{\Gamma_1,Q_4}-\Delta_{d}\Delta_{\Gamma_1,Q_2}^* \} +c.c.
\label{K-induc}
\end{equation}
and the second exists only in a finite magnetic field and is
\begin{equation}
\gamma_2 H_1S^z_{Q_2} \{ \Delta_{d}^*\Delta_{\Gamma_4,Q_4} + \Delta_{d}\Delta_{\Gamma_4,Q_2}^* \} + c.c.
\label{H-induc}
\end{equation}
where we have included $H_1$, the magnetic field along the
$(1,-1,0)$ direction. The experimental observation of a non-zero
$S^z_{Q_2}$ therefore leads to two possible types of PDW order. In
the time-reversal broken phase, the PDW order must belong to the
$\Gamma_1$ representation. In the time reversal-invariant phase,
the PDW order must belong to the $\Gamma_4$ representation. This
second possibility is most closely related to the $\pi$-triplet
staggered phase that has been found within a simple microscopic
description of CeCoIn$_5$ \cite{ape08}. Note that, in principle,
both the representations $\Gamma_1$ and $\Gamma_4$ will appear
simultaneously. However, it is reasonable to expect that one of
the two representations will give rise to the dominant order
parameter.\\

{\it Role of Vortices:} Now we examine the influence of the
vortices on the induced SDW order, in the mixed phase in a
magnetic field.  Prior to turning to the detailed analysis, we
present the two main results here: (i) The vortex cores of the two
PDW degrees of freedom $\Delta_{\Gamma_i,Q_2}$ and
$\Delta_{\Gamma_i,Q_4}$ can lie at
 different positions and also need not coincide with the $d$-wave
 vortex cores. We find  that there exist stable phases where this
 happens. These phases are defined by the relative displacements $\vtau_i$ of
 the PDW vortex cores from the $d$-wave vortex cores. In such
 phases, the $d$-wave vortex cores exhibit CDW order.
(ii) The SDW order leads to Bragg peaks that are determined by the
 reciprocal lattice vectors of the vortex lattice {\it and the
 displacements $\vtau_i$} (see Eq.~14).

For a detailed derivation of the above results,
we carry out the simplest realistic analysis. We
assume that the correlation length of the spin-density order is
much smaller than the coherence length of the superconducting
order. We take Eq.~\ref{H-induc} as the term driving the SDW order
(the same arguments can be applied if Eq.~\ref{K-induc} is used).
From this we obtain
\begin{equation}
S^z_{Q_2}(\vR)=\frac{\gamma_2H_1}{\alpha_s}[\Delta_{d}(\vR)^*\Delta_{\Gamma_4,Q_2}
(\vR)+\Delta_{d}(\vR)\Delta_{\Gamma_4,Q_4}^*(\vR)].
\end{equation}
The spatial dependence of the PDW and d-wave order parameter can
now be determined in the high-field limit for which the field $ H
$ may be considered uniform. From Eq.~\ref{d-free}, one finds that
the $d$-wave component yields an Abrikosov vortex lattice. Using
$z$ to represent the $(0,0,1)$-  and $x$ the $(1,1,0)$-direction,
the vortex lattice solution can be  given by
\begin{equation}
\Delta_d(\tilde{x},\tilde{z})=\Delta_{d0}\sum_n c_n
e^{iq(n-1/2)\tilde{x}}e^{-(\tilde{z}-z_n)^2/2}
\end{equation}
where $\tilde{x}=x/\epsilon$, $\tilde{z}=\epsilon z$, the vortex
lattice in the coordinates $\tilde{x},\tilde{z}$ has the basis
vectors $\va=(a,0)$ and $\vb=b(\cos\alpha,\sin\alpha)$
\cite{jam69}, $c_n=e^{i\pi\rho n^2}e^{-i\pi\rho(n+1)}$,
$q=2\pi/a$, $z_n=b\sin\alpha(n+1/2)$, $\rho=(b/a) \cos \alpha$,
and $\epsilon=[(\kappa-\kappa_c)/\kappa]^{1/4}$. The parameter
$\epsilon$ scales lengths in the $x$ and $z$ directions to take
the anisotropy into account.  This solution is an $n=0$ eigenstate
of the operator
$\tilde{\vD}^2=\tilde{D}_x^2+\tilde{D}_z^2=(-i\tilde{\vnabla}-2e\tilde{\vA})^2$
with eigenvalues $(2n+1)/l^2$ and $l^2=\Phi_0/(2\pi H)$ (
$n=0,1,2,..$ is the Landau level (LL) index). The macroscopic
degeneracy of the eigenstates of $\tilde{\vD}^2$ is exploited to
create the Abrikosov vortex lattice solutions and, at the same
time, plays a central part in constructing degenerate solutions
for the displaced vortex lattice
 ($\tilde{\phi}_n$) characterized by a vector $\vtau$:
$\tilde{\phi}_n(\br+\vtau)=e^{-i\tau_yx}\phi_n(\br+\vtau)$ with
$\phi_n({\br})$ being a vortex lattice solution in LL $n$. The states
$\tilde{\phi}_n$ and $\phi_n$ are degenerate eigenstates of the
operator $\tilde{\vD}^2$.

In order to determine the PDW vortex structure it suffices to
consider the linear equation for the PDW order parameter, which is
found by keeping both Eqs.~\ref{d-free} and \ref{PDWdc}, and by
setting $\beta_i=0$ in Eq.~\ref{d-free}. As a
technical simplification, we set
$(\kappa_1-|\kappa_2|)/\kappa_3=\kappa/(\kappa+\kappa_c)$ to
ensure that the $d$-wave order and the PDW order share the same
$\tilde{\vD}^2$ operator and hence have the same eigenstates (results without this simplification will be given elsewhere).
Minimization of the free energy yields the following for the two
degrees of freedom in the PDW order:
\begin{eqnarray}
\tilde{\Pi} \Delta_{\Gamma_4,Q_2}=&-\beta_{c1}|\Delta_d|^2\Delta_{\Gamma_4,Q_2}-\beta_{c2}\Delta_d^2\Delta_{\Gamma_4,Q_4}^*\nonumber\\
\tilde{\Pi}  \Delta_{\Gamma_4,Q_4}=&-\beta_{c1}|\Delta_d|^2\Delta_{\Gamma_4,Q_4}-\beta_{c2}\Delta_d^2\Delta_{\Gamma_4,Q_2}^*
\end{eqnarray}
with $ \tilde{\Pi} =
(\alpha+\sqrt{(\kappa_1-\kappa_2)(\kappa_1+\kappa_3)}\tilde{\vD}^2)
$. To solve these equations, we expand the PDW order in
eigenstates of the $\tilde{\vD}^2$ operator. At sufficiently high
fields, the PDW order will lie predominantly in the $n=0$
eigenstate for both $\Delta_{\Gamma_4,Q_2}$ and
$\Delta_{\Gamma_4,Q_4}$,  and we ignore the smaller higher $n$
contributions here. As mentioned above, these solutions are
degenerate, implying the use of two displacement vectors $\vtau_2$
and $\vtau_4$.  At the second order transition where the PDW order
appears, the vortex lattice structure is determined entirely by
the $d$-wave order parameter, so the only undetermined parameters
are $\vtau_2$ and $\vtau_4$. Solving the resulting linear equation
yields the result that the optimal PDW state is found by
minimizing
$\beta_{c1}\beta_A(\vtau_2)-|\beta_{c2}\tilde{\beta}(\vtau_2,\vtau_4)|$
with respect to $\vtau_2$ and $\vtau_4$ where \begin{eqnarray}
\beta_A(\vtau)=&&\sum_{{\bf G}}e^{-\frac{l^2G^2}{2}}e^{i{\bf
G}\cdot\vtau}\\\tilde{\beta}(\vtau_2,\vtau_4)=&&\sum_{{\bf
G}}e^{-\frac{l^2\tilde{G}^2}{2}}e^{i{\bf G}\cdot\vtau_4}
\label{beta-tau}
\end{eqnarray}
where ${\bf G}$ are the reciprocal lattice vectors of the vortex
lattice,
$\tilde{\bf G}={\bf G}+\frac{2\pi \vB}{\Phi_0}\times \vtau_2$ and
$\tilde{\beta}=0$ unless $\vtau_2+\vtau_4=\vT$, where $\vT$ is a
vortex lattice translation vector.
For $\beta_{c1}<0$ it follows immediately that $\vtau_2=\vtau_4=0$
while the solution for $\beta_{c1}>0$ requires a numerical
minimization to determine $\vtau_2$. The resulting configurations
are shown in Fig.~2, assuming that the $d$-wave order forms a
hexagonal vortex lattice.

Here we provide a description of the phases in Fig.~2. The phase
diagram depends upon $r=|\beta_{c2}/\beta_{c1}|$ and in all the
phases we can choose $\vtau_4=-\vtau_2$. We find that there are
four phases: in Phase 1 ($0\le r <0.07$), $\vtau_4=\gamma
(\va+\vb)$ and $\gamma$ evolves continuously from $1/3$ to $1/2$;
in Phase 2 ($0.07\le r <0.31$) $\gamma$ stays fixed at $1/2$ (Fig. 2
shows $\vtau_4=\va/2$ which is equivalent solution to
$\vtau_4=(\va+\vb)/2$); in Phase 3 ($0.31\le r <0.5$),
$\vtau_4=\gamma_2 \va$ where $\gamma_2$ changes continuously from
$1/2$ to $0$; finally in Phase 4 ($r>0.5$), $\vtau_4=0$. The
arguments of Ref.~\cite{agt08} imply that in Phases 1 through 3,
the $d$-wave vortex cores have charge density wave order at twice
the PDW wave-vectors.

\begin{figure}
\epsfxsize=3.0 in \center{\epsfbox{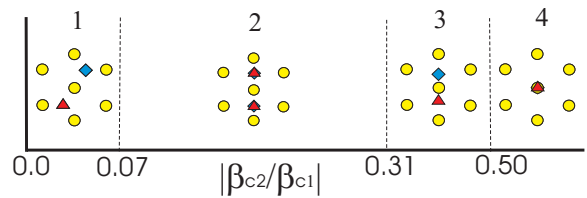}} \caption{Possible
vortex configurations for the PDW order. The yellow circles give
the zeroes of the $d$-wave order parameter, the blue diamonds give
the positions of the zeroes of $\Delta_{\Gamma_4,Q_4} $, and the
red triangles give the positions of the zeroes of
$\Delta_{\Gamma_4,Q_2}$.} \label{fig1}
\end{figure}

The solution of the vortex lattice problem for the PDW order
allows the SDW order to be determined which is particularly
important as neutron scattering measures the Fourier transform of
$S^z(\vR)$. Eq.(\ref{beta-tau}) yields the intriguing result that
the SDW order will exhibit Bragg peaks at $\vk$ positions that
depend upon $\vtau_2$ and $\vtau_4$:
\begin{equation}
\vk=\vQ_2+\vG+ \frac{2\pi { \vB}}{\Phi_0}\times \vtau
\end{equation}
where $\vG$ is a reciprocal lattice vector of the vortex lattice
and $\vtau$ is either $\vtau_2$ or $\vtau_4$. Consequently, the
relative position of the vortex cores of the PDW and $d$-wave
order can be retrieved from the position of the Bragg peaks in the
SDW order.



{\it Conclusions:} We have developed a phenomenological theory for
the $Q$-phase of CeCoIn$_5$ to identify the possible
symmetries for the PDW order.  This theory is used to determine
phases in which the PDW and $d$-wave vortex lattice are
relatively displaced, leading to CDW order in the $d$-wave vortex
cores. Interestingly, these structures can be probed by the
position of the SDW Bragg peaks.

D.F.A. is grateful for the hospitality of the Center for
Theoretical Studies at ETH Zurich. We thank Michel Kenzelmann for
useful discussions. We acknowledge financial support by Swiss
Nationalfonds and the NCCR MaNEP.


\end{document}